\documentclass[12pt,preprint]{aastex} 

\usepackage{natbib}

\begin{document}

\title{The effect of limited spatial resolution of stellar surface magnetic field maps on MHD wind and coronal X-ray emission models}

\author{ C. Garraffo\altaffilmark{1}, O. Cohen\altaffilmark{1}, J.J. Drake\altaffilmark{1}, C. Downs\altaffilmark{2}}

\altaffiltext{1}{Harvard-Smithsonian Center for Astrophysics, 60 Garden St. Cambridge, MA 02138}
\altaffiltext{2}{Institute for Astronomy, University of Hawaii at Manoa,
2680 Woodlawn Dr., Honolulu, HI 96822, USA }

\begin{abstract}

We study the influence of the spatial resolution on scales of $5\deg$ and smaller of solar surface magnetic field maps on
global magnetohydrodynamic solar wind models, and on a model of coronal heating
and X-ray emission.  We compare the solutions driven by a
low-resolution Wilcox Solar Observatory magnetic map, the same map with spatial resolution
artificially increased by a refinement algorithm, and a high-resolution Solar and Heliospheric Observatory Michelson Doppler Imager map.
We find that both the wind structure and the X-ray morphology are
affected by the fine-scale surface magnetic structure. Moreover, the X-ray morphology
is dominated by the closed loop structure between mixed polarities on smaller scales and shows
significant changes between high and low resolution maps.  We conclude that three-dimensional modeling of coronal X-ray emission has greater surface magnetic field spatial resolution requirements than wind modeling, and can be unreliable unless the dominant mixed polarity magnetic flux is properly resolved.

\end{abstract}

\keywords{stars: magnetic field - stars: coronae -}

\section{INTRODUCTION}
\label{sec:Intro}

Stellar magnetic fields play an important role in many astrophysical
phenomena.  In particular, they are known to dominate the structure and X-ray morphology of solar and stellar coronae \citep[e.g.][]{Aschwanden04,Gudel07}, of the solar wind \citep{parker58} and, presumably, its stellar wind analogs.  Thanks to new-generation high-resolution spectropolarimeters, such as ESPaDOns 
 \citep{manset}, we are 
able to study the large-scale magnetic structure of stars
and to extract maps of stellar surface magnetic fields (hereafter magnetograms).  These magnetograms can be used as the boundary condition for extrapolations of the extended magnetic-field distribution in the corona and the stellar wind.  However, the spatial resolution of stellar magnetic field maps, being intrinsically limited by the precision of observational techniques, is much lower than those that can be obtained for the Sun, and no comparable high-resolution stellar maps are available.  

At present, the most detailed and reliable available stellar surface magnetic field
maps are those constructed using the Zeeman-Doppler
Imaging technique (ZDI) (\citet{DonatiCollierCameron97,Donati99}; see \citealt{Donati09} for a review).  
Doppler imaging in the stellar context is an indirect imaging technique that uses rotational phase-dependent deformations in spectral-line
profiles to map starspots \citep{Vogt83}.  Both Doppler imaging
and spot modeling have been widely applied to map the surface of a
number of solar-like stars
\citep[e.g.]{Strassmeier98,Barnes98,Jarvinen05}.  More recently, ZDI was developed as an extension to Doppler imaging by introducing spectropolarimetric
observations of lines.  
This technique takes into account two of the five
components of the Stokes vector (Stokes I and Stokes V) as well as
rotation, and decomposes the magnetic field into a
poloidal and a toroidal component.  ZDI maps have so far been constructed for
a wide range of late-type stars, including solar analogs, K and M dwarfs, and pre-main sequence stars \citep[e.g.][]{Donati03,Marsden06,Catala07,Donati08,DonatiMorin08,Donati10}.
ZDI involves
Least-Square Deconvolution (LSD),
which assumes
self-similarity and linear addition of line blends, and its range of validity was not clear until further analyzed
by \cite{Kochu}.  After a detailed study of each Stokes
component validity range, \cite{Kochu}
concluded that the LSD profiles are reasonably robust for the
circular polarization spectra (Stokes V) in the field strength range 0-2 kG.  For stars with surface magnetic fields within this range one can assume the ZDI technique is, at least in principle, reliable.              
 
Since late 1960s great progress has been made in the understanding and modeling of
the coronal structure of the sun and stars. \citet{parker58} provided the first hydrodynamic solution for the distribution of the solar wind as a function of radius.  The first model that took into account both the solar wind and the coronal magnetic field was introduced by \citet{Pneuman71}. This study of the gas-magnetic field interaction in
the solar corona laid the foundations for numerous
simulations of the solar coronal steady-state conditions.  
A common way to extrapolate the coronal structure from the magnetic
fields in the photosphere is
the so called Potential Field Source Surface method \citep{Altschuler69}.  This method has been applied to ZDI maps for several stellar systems \citep[e.g.][]{Jardine02a, Jardine02b, Hussain02, McIvor03, Hussain07, Donati08}.  However, this approximation
is based on strong assumptions about the boundary conditions and force balance in the system.  In
particular, it assumes that there are no currents in the system
and, thus, that the magnetic field can be described as a gradient of a scalar
potential. The three dimensional magnetic structure of the corona is
then obtained by solving the Laplace equation for this scalar
potential, using the magnetogram as a boundary condition together with an outer
boundary condition, where it is assumed that the magnetic field is purely radial.  The idea behind this is
that the solar wind's dynamic pressure overcomes the magnetic pressure of
the stellar magnetic field at some height, opening the magnetic field
into the heliosphere.  It is not clear, however, at what distance
this outer boundary should be set and whether this boundary is
spherical \citep{Riley06, Gilbert07}. 

 ZDI magnetograms are only able to reproduce the large
 -scale surface magnetic structure of stars, therefore they lack
information.
 Some of the effects of missing magnetic flux on the
  stellar coronae modeling have been studied by \cite{Johnstone10} and \cite{Arzoumanian11} using the
  Potential Field Approximation method. 
  They conclude that
  the finite spatial resolution limitation is most relevant for highly
  complex fields, though missing flux in dark spots can lead to overestimating open flux, which might adversely affect stellar wind models.
  
 A more realistic approach for simulating a stellar corona would be to use MHD models \cite[e.g.]{Usmanov93, Mikic99,Wu99,
Suess99,Groth00,Usmanov03,Roussev03}.  In particular, \cite{cohen07}
 developed a semi-empirical MHD model for the solar corona.
An advantage of MHD models over the
Potential Field Approximation is that they provide a steady state,
non-potential solution that includes the complete set of parameters of the
system, and not just the magnetic field.

X-ray emission of solar-like stars is mostly generated by the hot
plasma confined in close coronal loops with footpoints in the photosphere,
presumably with a smaller contribution from flaring activity
\citep[e.g.][]{Vaiana78,Gudel97,Drake00a,Testa04}.  The sizes of these emitting loops likely depend
strongly on the properties of each star (e.g. rotation rate, which governs the activity level,  and spectral type)
and have been investigated using X-ray spectrophotometry and simple models 
\citep[e.g.][]{Giampapa85, Stern86, Schrijver89, Giampapa96, Gudel97b,
  Preibisch97, Sciortino99, Stern86, Giampapa96, Maggio97, Ventura98,
  Sciortino99}, 
  coronal density measurements based on X-ray spectroscopy
\citep[e.g.][]{Ness02, Testa04}, rotation modulation
\citep[e.g.][]{Drake94, Gudel95a, Gudel95c,Brickhouse98,Brickhouse01, Marino03a, Hussain05}, and eclipse mapping
techniques \citep{Schmitt93,Schmitt99, Favata99, Gudel03a, Favata05, Mullan06,
Hussain07}. 
While these techniques help to locate the emitting regions and constrain the source sizes, 
the X-ray luminosity depends crucially on the heating
rate of the emitting loop and is therefore strongly dependent on the
heating model used 
\citep{Schrijver02}.   Although there are many coronal heating models in the
literature \citep[e.g.][]{Schrijver02,Schrijver04, Abbett07, Mok08}, the low coronal heating mechanisms are
still far from understood.  Recently, a new feature of the {\sc BATS-R-US} model, called the {\it Low
  Corona} module, has been developed by \cite{Downs10}
(described below in Section~\ref{sec:Model}).  This module reproduces
the X-ray morphology of the low corona of the sun by computing the MHD
magnetic field structure and filling the closed field with hot
emitting plasma, assuming a simple empirical heating
model.  It has been shown to 
reproduce successfully the salient features of the observed solar coronal X-ray structure \citep{Downs10}.    

In this paper we study the relevance
of spatial resolution of the surface magnetic field maps on these MHD
wind and X-ray models and, in consequence, how
reliable the results of stellar coronal and wind models based on
available low-resolution stellar magnetograms are likely to be.   We use the solar case as a
laboratory, utilizing the obvious benefit of being able to compare model predictions with high quality observations of the solar wind and the X-ray Sun.  
Using an MHD code we simulate the three-dimensional solar wind 
structure and coronal X-ray emission morphology, and we investigate the influence on the solutions of small-scale structure in the input magnetogram.  
We also develop a method to model the 
surface magnetic
structure of a low-resolution magnetogram as well as an algorithm to artificially increase its
resolution.  We apply this 
to a low-resolution solar magnetogram and compare the MHD wind and corona 
solutions driven by this artificially refined magnetogram with those 
computed for the low-resolution magnetogram and for a contemporaneous high-resolution map obtained by the Solar and Heliospheric Observatory (SoHO) Michelson Doppler Imager (MDI).\footnote{http://sun.stanford.edu}

The numerical methods are described in Section~\ref{sec:Methods}.  In
Section~\ref{sec:Results} we present our results.  We discuss
our main findings and their implications in
Section~\ref{sec:Discussion}. The results
are summarized in Section~\ref{sec:Conclusions}.


\section{NUMERICAL SIMULATION}
\label{sec:Methods}

\subsection{MHD model} 
\label{sec:Model}

We simulate the solar corona using the BATS-R-US global MHD model,
which was originally developed by \cite{powell99,toth05}, and was
later adapted to study the solar coronae by \cite{cohen07}. It was
also used to simulate stellar coronae (see e.g.,
\cite{cohen10a,cohen10b}). The model provides a self consistent,
steady-state solution for the solar corona and its wind structure by
solving the set of conservation laws for the mass, momentum, magnetic
induction, and energy of the coronal plasma.  The solar wind model is
driven by the surface magnetic field data (magnetograms), while the
solar wind powering is specified semi-empirically using the empirical
relation between the solar wind terminal speed and the magnetic flux
tube expansion factor, defined as $f_s =
(R_{\odot}/R_{ss})^2[B(R_\odot)/B(R_{ss})]$, commonly known as the
Wang-Sheeley-Arge (WSA) model \citep{wang90,argepizzo00}.  Here $ B $
is the field strength and the subscript
$ss$ stands for {\it source surface}.  The WSA model has been successful in predicting the solar wind speed at 1~AU, but it has some limitations which are described below.

The algorithm to obtain a steady-state solution is as follows (we
refer the reader to \cite{cohen07} for a full description of
the model). First, the three-dimensional potential field is calculated
based on the magnetogram data and assuming a source surface at some
distance ($2.5R_\odot$ for solar simulations presented here). The
calculation is done after the magnetogram has been converted to a set
of spherical harmonics coefficients with an order of equals to the
original map resolution, so that artifact such as ``ringing'' effect do not appear in the solution \citep{toth11}. No interpolation of the polar field is done beyond the interpolation done by the observatories themselves.  Once the potential field is set, the angular distribution of the terminal wind speed is calculated using the WSA model. This distribution is then used to specify an angular distribution of the polytropic index, $\gamma$ assuming a conservation of the Bernoulli Integral along the particular magnetic field line. Finally, an energy source term is specified and constrained by this surface distribution of $\gamma$ and then the MHD solution is self-consistently converging to a steady-state solar wind solution. While the WSA model by itself provides the solution for the wind speed only (and the magnetic field polarity), the algorithm described here provides the full MHD solution, which is constrained by the WSA model wind speed.

For the X-ray emission simulations we use the {\it Low Corona} module, which has been shown able to reproduce quite accurately  the general X-ray morphology of 
the solar corona \citep{Downs10}. This module
calculates the coronal EUV and soft X-ray synthetic emission based on the
surface magnetic field, using an MHD model, which includes an empirical coronal heating model.  In particular, it takes into account
non-MHD thermodynamic terms by adding them to the governing
MHD energy equation in the following way
$$ \frac{\partial E}{\partial t}+\nabla\cdot\left( 
\mathbf{F}_{MHD}+\mathbf{F}_{c}  \right)=Q_{MHD}+Q_r+Q_h.$$
where the subscript $MHD$ refers to the MHD solution with no heat sources, while   
$F_c$,  $Q_r$ , $Q_h$ represent the
electron heat conduction, the radiative losses and the heating term
respectively (see \citet{Downs10} and references therein for details).
Solving for this equation, the {\it Low Corona} model then fills the flux
tubes of an MHD solution according to the chromospheric base density and the heating and
cooling, in a self-consistent physical manner and with two degrees of
freedom corresponding to the magnetic field and the loop height.
The model provides us with a distribution of the magnetically confined X-ray
emitting gas which we can compare with observations. 
We use an exponential empirical coronal heating model which
gives the best match to {\it Yohkoh} Soft X-ray Telescope
data\footnote{http://ylstone.physics.montana.edu}. We then compare the resulting X-ray morphologies.

While the approach described above has been successful in reproducing
the solar wind conditions \citep{cohen08}, it is important to mention
its limitations. In particular, it is pretty clear that the WSA model
is sensitive to magnetogram resolution \citep{Jian11,
  Mcgregor11, Riley12}, since the resolution determines the magnetic field
mapping, and the structure of the different flux-tubes. In the
simulations presented here, we use high-resolution solar MDI
magnetogram with a set of spherical harmonic coefficients of the order
of n=90, while the WSO maps and the artificially refined maps are of
the order of n=73.  We do not attempt to make the WSO data look like
the MDI one, but rather to investigate how the changes in the data
itself,  {\it while using the same magnetogram and MHD grid resolution}, affect the wind and X-ray solutions. The reference case for the MDI data is simply to show how modeling based on actual high-resolution data looks like. Moreover, the original goal of the model in the solar context is to predict the solar wind conditions at 1~AU, and for such a prediction, the issues related to the WSA model are of great importance. However here, we only study how changing the data itself to include more structure, while keeping the magnetogram resolution the same, affect the global wind structure, as well as the X-ray solution. In the last stage of specifying the energy source term, the potential field (or magnetogram) grid is interpolated to the MHD grid. In the simulations presented here, the smallest grid cell near the solar surface is of a size of $\Delta x=0.015R_\odot$, which is about 5 degrees, similar to the magnetogram grid size.

\subsection{THE STELLAR MAGNETIC FIELD}
\label{sec:StellarField}

\subsubsection{Magnetograms}
\label{sec:Magnetograms}

In order to understand the effect of spatial
resolution of the stellar surface magnetic fields on the stellar wind 
and X-ray emission models,
we compare simulations driven by both low and high-resolution magnetograms. While solar observations have excellent spatial 
resolution, only low-resolution magnetograms are
available for stars. For this reason we consider three solar
magnetograms for our analysis: a low-resolution solar magnetogram obtained from the Wilcox Solar Observatory (WSO)\footnote{http://wso.stanford.edu}; the
 same magnetogram with spatial resolution artificially increased by the
 algorithm described below in Section~\ref{sec:HREM}; and a
 high-resolution MDI map\footnote{http://sun.stanford.edu/synop/};.  By comparing the different solutions we are able to establish
 how the solution depends on the spatial resolution of the
 magnetogram. 
The spatial resolution of ZDI magnetograms strongly depends on
  the projected equatorial velocity of the star. Typically, the
  resolving power at the equator can range from a few degrees for
  some very fast rotators such as the K1 dwarf AB Doradus \citep{Donati97, Donati99}, to $50^\circ$
  for slow rotators like the K2 dwarf HD 189733 \citep{Fares10}.  In the best
  stellar cases the
  resolution is comparable to that of the WSO solar maps but still
  very far from the one achieved with MDI observations, for which the
  resolution is better than a tenth of a degree on the sun. Hereafter, by
  ``high-resolution'' we will refer to equatorial resolving power of less
  than one degree.

\subsubsection{Automatic Magnetogram Modeling Algorithm}
\label{sec:AMMA}

To model the low-resolution magnetograms we have developed an algorithm that automatically
reproduces an image through a collection of three-dimensional Gaussian
functions on a
sphere. Each magnetic pole is replaced by a function of appropriate amplitude and dispersions (in
principle different in the $\theta$ and $\phi$ directions).
Since
Gaussian functions do not form a complete basis, it is not possible to
expand a data set on them as one would do with any set of the, so called, special
functions. However, our interest in Gaussians relies on the fact that
these are localized functions and, therefore, they better represent
some physical phenomena such as magnetic poles.
Our method is based on the Expectation Maximization algorithm approach
(\cite{Dempster77,Redner84,Wu86}), which is a numerical method that estimates the parameters
of a set of probability density functions that most likely have generated a given set of data
points. Starting from some initial set of values for the parameters,
this algorithm iteratively finds a better set of them by maximizing
the likelihood function. However, this algorithm is only suitable for a definite
positive set of data points and for a collection of positive definite
Gaussian functions on the plane. 
We modify the Expectation Maximization algorithm in order to reproduce images with both positive and
negatives intensity values, with periodic boundary conditions in
one of the variables and with a spherical base manifold. 
Since the convergence of these algorithms strongly depends on
the initial conditions, we also develop a method that provides a good
first order estimation of Gaussian parameters. At the 
local maximum of each map, the later places a Gaussian function with amplitude equal to the intensity of the map at
that point and a dispersion of a certain fixed value that depends on
the scale of the magnetogram. For solar low-resolution magnetograms we
use 360/15 degrees.

\subsubsection{High Resolution Extrapolation Method}
\label{sec:HREM}

When looking at surface magnetic fields with the complexity of that of the Sun using low-resolution technology we miss
the underlying fine structure.  A few large magnetic poles are
usually enough to model the low-resolution data, although the true magnetic structure 
is less simple.  Based on the solar case, one
can safely assume that a magnetic area of a given polarity is
produced by many small magnetic poles rather that just a big
one. So, in order to ``increase'' the resolution of a stellar surface
magnetic map  represented by
a few large magnetic poles (i.e. to add small-scale structure to it), a first step would be to find a larger distribution 
of poles that will reproduce the original low-resolution map. In other words, we wish to
introduce a more realistic spatial distribution of the data while retaining the characteristics of the low
resolution image.
Once we have obtained a model of 
fine structure that matches the low-resolution data, we can examine how this new distribution would actually look when observed with
a higher resolution observational technique. 

For this purpose we have developed the {\it High Resolution
  Extrapolation Method} (HREM) that replaces each magnetic pole in
the original automatically modeled map by a distribution of twenty-one smaller magnetic poles.
One is placed at the center of the original magnetic pole while the remaining twenty are distributed over
two ellipses, also centered on the original magnetic pole, whose parameters   
are chosen to minimize the difference between the integral of the
original Gaussian and the sum of the integrals of the new distribution.   The
reason behind this is that we wish the flux to be conserved.
We can now express our original function as a sum of twenty-one new
Gaussians distributed around the original one,
\begin{eqnarray}
 \lefteqn{G_{A,\phi_0, \theta_0, \sigma_{\phi}, \sigma_{\theta}}(\phi,\theta
  )  \approx g \, A \,  \big[G\Phi_{\phi_0, d \, \sigma_{\phi} } \, G\Theta_{
  \theta_0, d \, \sigma_{\theta}}} \nonumber\\
  &&+  \, \,  e^{-2/5} \big( G\Phi_{\phi_0 \pm b \,
  \sigma_{\phi}, d \, \sigma_{\phi}  }
  \, G\theta_{\theta_0, d \, \sigma_{\theta} }
+ G\Phi_{\phi_0, d \,  \sigma_{\phi}} \,  G\theta_{\theta_0  \pm b \,
   \sigma_{\theta}, d \,  \sigma_{\theta}} \nonumber \\
   && + \,  G\Phi_{\phi_0 \pm cos(\pi/4) \, b \,
  \sigma_{\phi}, d \, \sigma_{\phi}  } \, 
  G\theta_{\theta_0  \pm cos(\pi/4) \,  b \, \sigma_{\theta},
  d \, \sigma_{\theta} } \big)  \nonumber \\
 &&+ \,  e^{-8/5} \big( G\Phi_{\phi_0 \pm c \, \sigma_{\phi}, d \,  \sigma_{\phi} }
  \, G\theta_{\theta_0, d \, \sigma_{\theta} } + G\Phi_{\phi_0, d \, \sigma_{\phi} } \,  G\theta_{\theta_0  \pm c \,
  \sigma_{\theta}, d \,  \sigma_{\theta}}  \nonumber \\
 && + \, G\Phi_{\phi_0 \pm cos(\pi/6) \, c \,
  \sigma_{\phi}, d \,  \sigma_{\phi} } \, 
  G\theta_{\theta_0  \pm sin(\pi/6) \,  c \, \sigma_{\theta},
  d \, \sigma_{\theta}}  \nonumber \\
&& +\, G\Phi_{\phi_0 \pm cos(\pi/3) \, c \,
  \sigma_{\phi}, d \,  \sigma_{\phi} } \, 
  G\theta_{\theta_0  \pm sin(\pi/3) \,  c \, \sigma_{\theta}, d \,
  \sigma_{\theta}} \big) \big] \nonumber
\end{eqnarray}
where $G_{a \pm b} = G_{a + b} + G_{a - b}$.
Here we have used the following compact notation,
\begin{eqnarray}
 G\Phi_{\phi_0, \sigma_{\phi}} &=&  exp{\big(-\frac{((\phi-\phi_0) \,
  cos(\theta)^2}{\sigma^2_{\phi}}\big)} \label{cos}  \nonumber \\
 G\theta_{\theta_0, \sigma_{\theta}} &=&  exp{\big(-\frac{(\theta-\theta_0)^2}{\sigma^2_{\theta}}
  \big)} \nonumber
\end{eqnarray}
so that the original Gaussian reads 
$$\, G_{A,\phi_0, \theta_0, \sigma_{\phi}, \sigma_{\theta} }(\phi, \theta
) =A \,  G\Phi_{\phi_0, \sigma_{\phi}}(\phi) \, G\theta_{ \theta_0, \sigma_{\theta}}(\theta). $$
So far we have a general expression for the original
magnetic pole with four free parameters, $g, b, c, d$. By means of minimizing 
the modulo of the flux difference, characterized by the
following function,
\begin{eqnarray}
 \lefteqn{\int^{\theta_0+3 \sigma_{\theta}}_{\theta_0-3\sigma_{\theta}}\int^{\phi_0+3 \sigma_{\phi}}_{\phi_0-3 \sigma_{\phi}} G_{A,\phi_0, \theta_0, \sigma_{\phi}, \sigma_{\theta} }(\phi,
\theta) d\phi \, d\theta} \nonumber\\
&& - \int^{\theta_0+3\sigma_{\theta}}_{\theta_0-3\sigma_{\theta}}\int^{\phi_0+3\sigma_{\phi}}_{\phi_0-3 \sigma_{\phi}}
 g \, A \,  \big[G\Phi_{\phi_0, d \, \sigma_{\phi} } \, G\Theta_{
  \theta_0, d \, \sigma_{\theta}} \nonumber\\
  &&+  \, \,  e^{-2/5} \big( G\Phi_{\phi_0 \pm b \,
  \sigma_{\phi}, d \, \sigma_{\phi}  }
  \, G\theta_{\theta_0, d \, \sigma_{\theta} }
+ G\Phi_{\phi_0, d \,  \sigma_{\phi}} \,  G\theta_{\theta_0  \pm b \,
   \sigma_{\theta}, d \,  \sigma_{\theta}} \nonumber \\
   && + \,  G\Phi_{\phi_0 \pm cos(\pi/4) \, b \,
  \sigma_{\phi}, d \, \sigma_{\phi}  } \, 
  G\theta_{\theta_0  \pm cos(\pi/4) \,  b \, \sigma_{\theta},
  d \, \sigma_{\theta} } \big)  \nonumber \\
 &&+ \,  e^{-8/5} \big( G\Phi_{\phi_0 \pm c \, \sigma_{\phi}, d \,  \sigma_{\phi} }
  \, G\theta_{\theta_0, d \, \sigma_{\theta} } + G\Phi_{\phi_0, d \, \sigma_{\phi} } \,  G\theta_{\theta_0  \pm c \,
  \sigma_{\theta}, d \,  \sigma_{\theta}}  \nonumber \\
 && + \, G\Phi_{\phi_0 \pm cos(\pi/6) \, c \,
  \sigma_{\phi}, d \,  \sigma_{\phi} } \, 
  G\theta_{\theta_0  \pm sin(\pi/6) \,  c \, \sigma_{\theta},
  d \, \sigma_{\theta}}  \nonumber \\
&& +\, G\Phi_{\phi_0 \pm cos(\pi/3) \, c \,
  \sigma_{\phi}, d \,  \sigma_{\phi} } \, 
  G\theta_{\theta_0  \pm sin(\pi/3) \,  c \, \sigma_{\theta}, d \,
  \sigma_{\theta}} \big) \big]  d\phi \, d\theta   \nonumber
\end{eqnarray}
we determine these parameters to take the values $ g = 0.3872 \sqrt{2
  \pi/5 }, \,  b = 0.6831 , \,  c = 1.2294 \, \, \mathrm{and} \, \, d =
  0.5121 $ \footnote{
Given the spherical symmetry of the background geometry that we are working on (the surface of the star), we need to include a factor $ cos(\theta_0) $, that will convert the angles into distances to properly represent the three-dimensional Gaussian function (as shown in equation~\ref{cos}).  Before performing the minimization procedure, we have absorbed the factor $cos(\theta_{0})$  within $\sigma_{\theta_0} $ by redefining it as  $  \sigma_{\phi}=\sigma_{\phi}/cos(\phi)$.  Once we have done this, we are able to
minimize the Gaussian as if it were on the plane. However, when
  applying our refinement method, and because we are distributing the
  new Gaussian over two ellipses around the original pole's position,
  some new magnetic poles will be at a different latitude ($\theta_{0
  \, new}$ ) than that of the original ($ \theta_0$). In these cases, the right factor to include in the new Gaussian function would be $ cos(\theta_{0 \, new})$ rather than $cos(\theta_0)$. In order to account for this, when applying our refinement method we multiply every new 
$ \sigma_{\phi}$ (which has absorbed a factor $ \frac{1}{cos(\theta)}$) by  $ \frac{cos(\theta_0)}{cos(\theta_{0 \, new})}$.
This introduces a slight error, as compared with the same
prescription carried out on the plane, due to the
fact that we are not using exactly the $\sigma_{\theta_0} $ that assures the minimum
flux difference (the reason for this is, in the end, that
here the angles are not distances until we introduce the $cos(\theta_{0})$ factor). This spherical geometry is also responsible of the
larger flux difference for magnetic poles of higher latitudes. However, we believe
this is the right geometrical procedure and the error introduced by it is negligible.}.

\begin{figure*}
\centering
\includegraphics[width=5in]{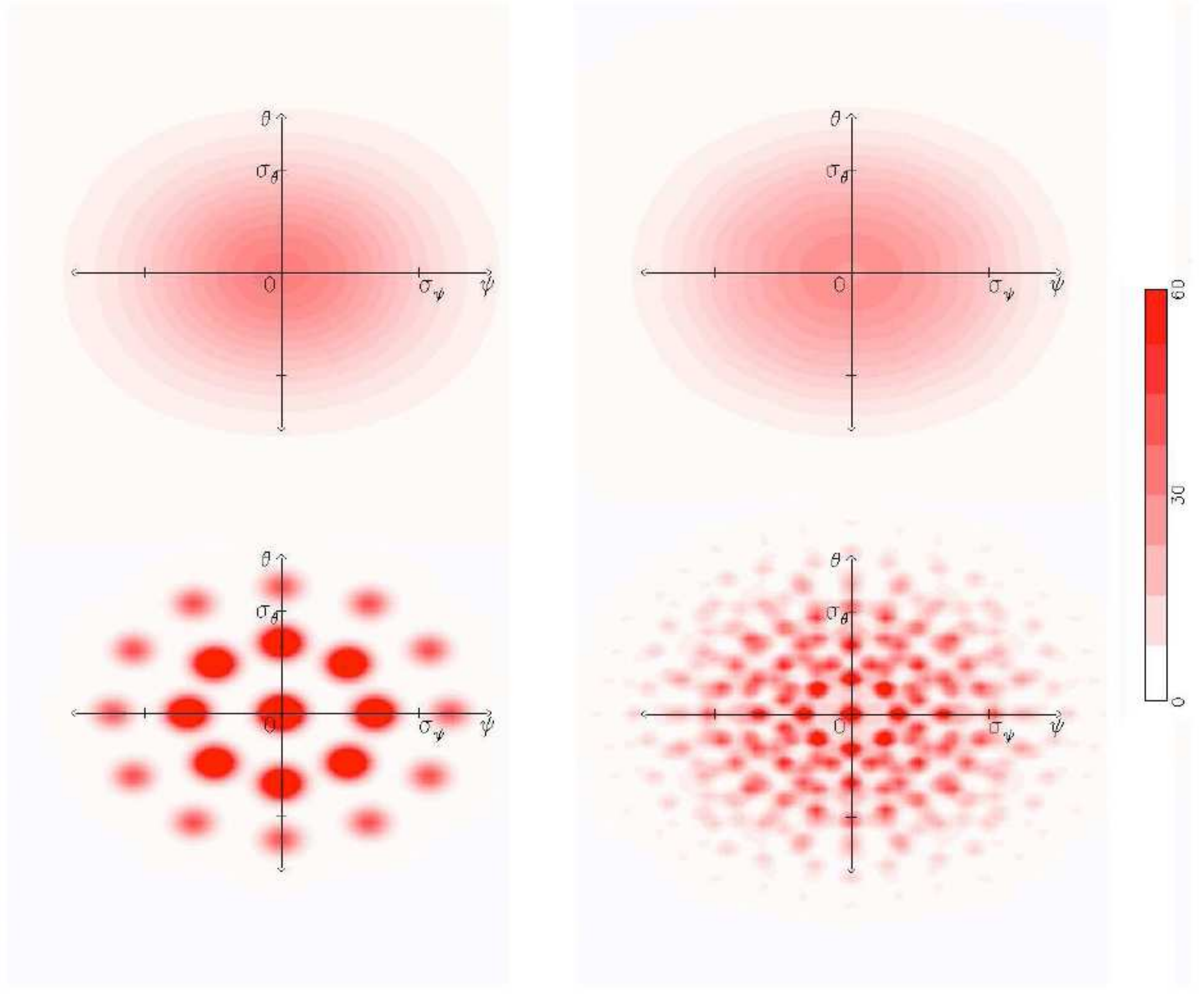}
\caption{Top: map generated by a single two-dimensional Gaussian centered at the origin, with dispersions $\sigma_{\Phi}=20$ and
  $\sigma_{\Theta}=15$ (left), and the corresponding maps resulting
  from applying the high resolution extrapolation method twice
  (right).  Bottom: Artificially
  refined maps taking a=4 for one (left) and two (right) iterations of the method.  Shading is based on a linear scale illustrated at right.}
\label{fig:PolesPosition}
\end{figure*}

Each time this method is applied it provides 21 new magnetic poles per
original one. Therefore, the number of iterations determines the number of new
poles and, in consequence, the amount of fine structure being
introduced.   
The top left plot of Figure~\ref{fig:PolesPosition} shows a single
Gaussian function with dispersions of $
\sigma_{\theta}=15$ and $\sigma_{\phi}=20$ degrees, which are typical
values for magnetic poles in these kind of maps. The top right plot of the same figure
shows the result of applying the refining algorithm twice to it. This map is generated by a distribution of 441 new Gaussian functions (21
per iteration for each pole).   
It is clear from the figure that we are able to reproduce
very accurately the original Gaussian.

So far we have discussed a method to
find one of the possible magnetic maps with fine-structure that in low-resolution
would look like the original map.
The next step is to find how the high-resolution version of this new map would look. This is straightforward if we keep in mind that, for
large enough integration intervals so that the functions die off
almost completely inside the integration domain, the following holds 
\begin{eqnarray}
\label{property}
{\cal{F}}=\int G_{A,\phi_0, \theta_0, \sigma_{\phi}, \sigma_{\theta} }(\phi, \theta) d\phi \, d\theta \approx \nonumber \\
a^2 \int G_{A,\phi_0, \theta_0, \frac{\sigma_{\phi}}{a},\frac{ \sigma_{\theta}}{a} }(\phi, \theta)  d\phi \, d\theta 
\end{eqnarray}
where $ \cal{F} $ is the magnetic flux associated with $G_{A,\phi_0, \theta_0, \sigma_{\phi}, \sigma_{\theta} }(\phi, \theta)$.
This property allows us to better resolve the magnetic poles by increasing $a$,
while conserving the flux of each one of them by decreasing $\sigma_{\phi,\theta}$ commensurately.  
Notice that $a$ is a choice that will determine how well resolved the
new fine structure given by this method is.  In this sense, $a$ is a
scale parameter.  The bottom panel of
Figure~\ref{fig:PolesPosition} shows the result of increasing the
parameter $a$ to $4$ for the refined maps after one and two iterations.

 
\section{RESULTS}
\label{sec:Results}

We simulate the solar wind and the X-ray coronal morphology for the
 low-resolution solar magnetogram, the same magnetogram with artificially
 increased spatial resolution, and the high-resolution MDI
 magnetogram.  For the solar wind we perform MHD simulations using the BATS-R-US code and for the X-ray coronal we use the new {\it Low Corona}
 module of the same code. 

By applying the HREM to the WSO low-resolution
magnetogram we get an artificially increased spatial resolution map
(hereafter processed map). The best agreement with
the scale of the MDI magnetograms is reached with two iterations and for
$a=10$.  

\begin{figure*}
\centering
\includegraphics[width=\textwidth]{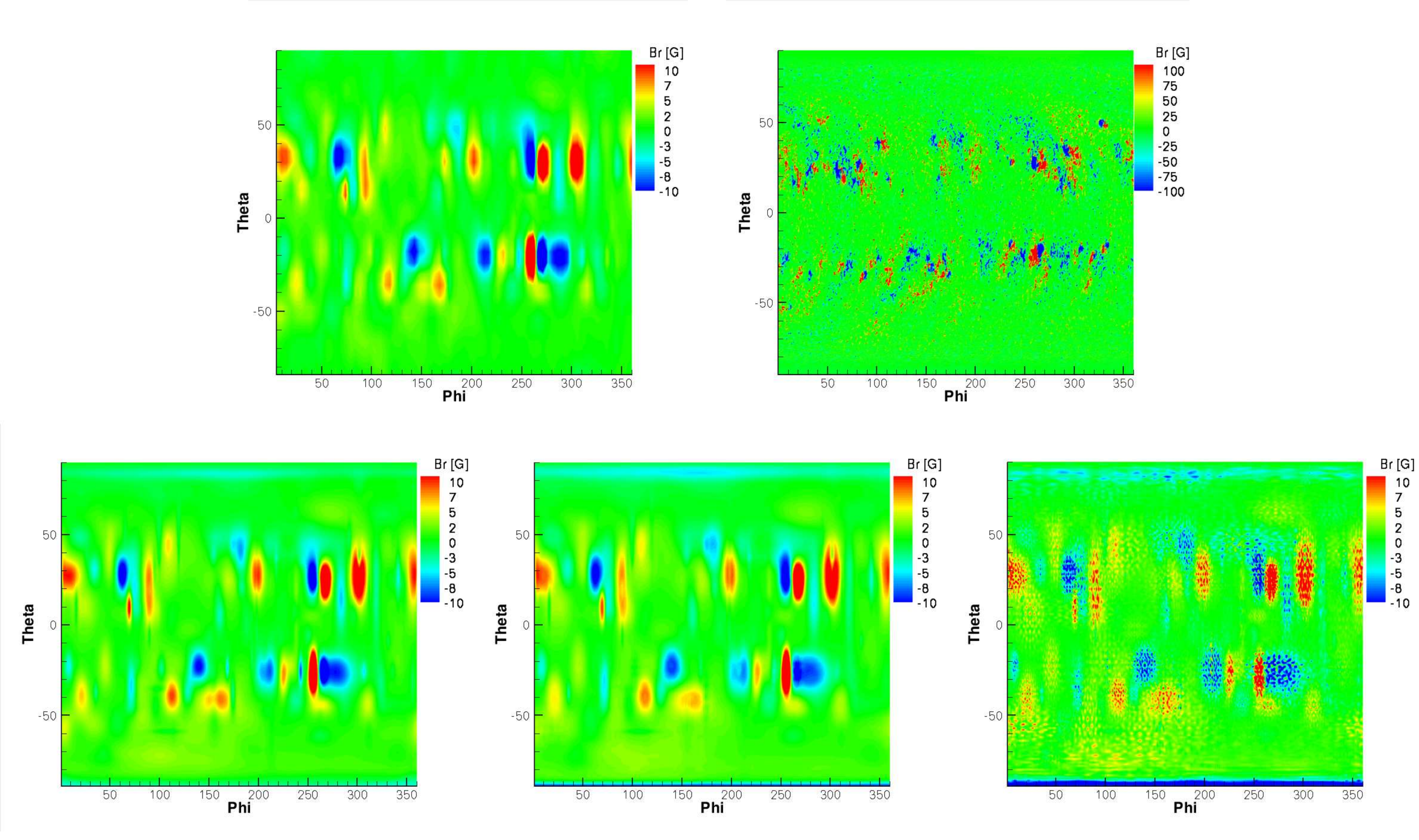}
\caption{Top: low- and high-resolution observations of the solar
  magnetic field during Carrington Rotation 1962 (WSO and MDI
  data).  Bottom: Gaussian representation of the WSO low resolution map
  (left), processed map with $a=1$ (middle), and the processed map with
  $a=10$ (right).  }
\label{fig:magnetograms}
\end{figure*}

The top plots of Figure~\ref{fig:magnetograms} correspond to the WSO
and MDI real data magnetograms. While the bottom panel shows the WSO
modeled magnetogram (left), the processed map that results from
applying two iterations of the HREM to it (middle), and the same map
after boosting the scale parameter $a$ from $1$ to $10$ (right).  At each
iteration the number of poles is increased 21 times, so two
iterations result in 441 poles per original one.
Nevertheless, the difference in the total
flux is only of 1.7 \% --- excellent agreement considering the
large number of new poles that are being introduced.  The
plots corresponding to the original map and to the high-resolution extrapolation
also match qualitatively,
as shown in Figure~\ref{fig:magnetograms} (two bottom left plots).  Choosing $a=10$ and applying property (\ref{property}) to all the new poles, we are
able to better resolve the new magnetogram, thus bringing to light the fine
structure lying behind (right bottom plot of Figure~\ref{fig:magnetograms}). This, of
 course, introduces an error (again, mainly due to the spherical
 symmetry of the base manifold) which increases with $a$. In our case, where
 $a=10$, the total flux difference increments to 2.8~\%.  In
 consideration of the fairly large uncertainties involved in measuring
 stellar surface magnetic field strengths we view this is as
 acceptable taking into account that, since we are dealing with
 three-dimensional Gaussians, each amplitude is boosted by $a^2=100$.   
Based on the good 
agreement between the original magnetogram and the new one we construct using
this prescription (bottom plots of Figure~\ref{fig:magnetograms}), together
with the extremely good conservation of flux, we conclude that the refinement method is robust. 
It is worth to point out that, as we have already discussed, the new enhanced map is only one of the
possible high-resolution maps that would recreate the observed WSO
one when smeared, as there is information missing in the WSO
map. Therefore we do not expect the processed map to reproduce the MDI
map. Our attempt is, rather, to study the
effects of introducing fine structure in a reasonable
way. By this we mean structure of the same scale of the MDI map and that would
recreate the WSO when decreasing its resolution while conserving the original
flux, which is different from the MDI flux by two orders of magnitude. 

The MHD solar wind solutions and the X-ray morphology simulations are
 shown in Figure~\ref{fig:WindSolutions} and Figure~\ref{fig:SXR_all} respectively.

\subsection{Solar Wind Structure}

\begin{figure*}
\centering
\includegraphics[width=\textwidth]{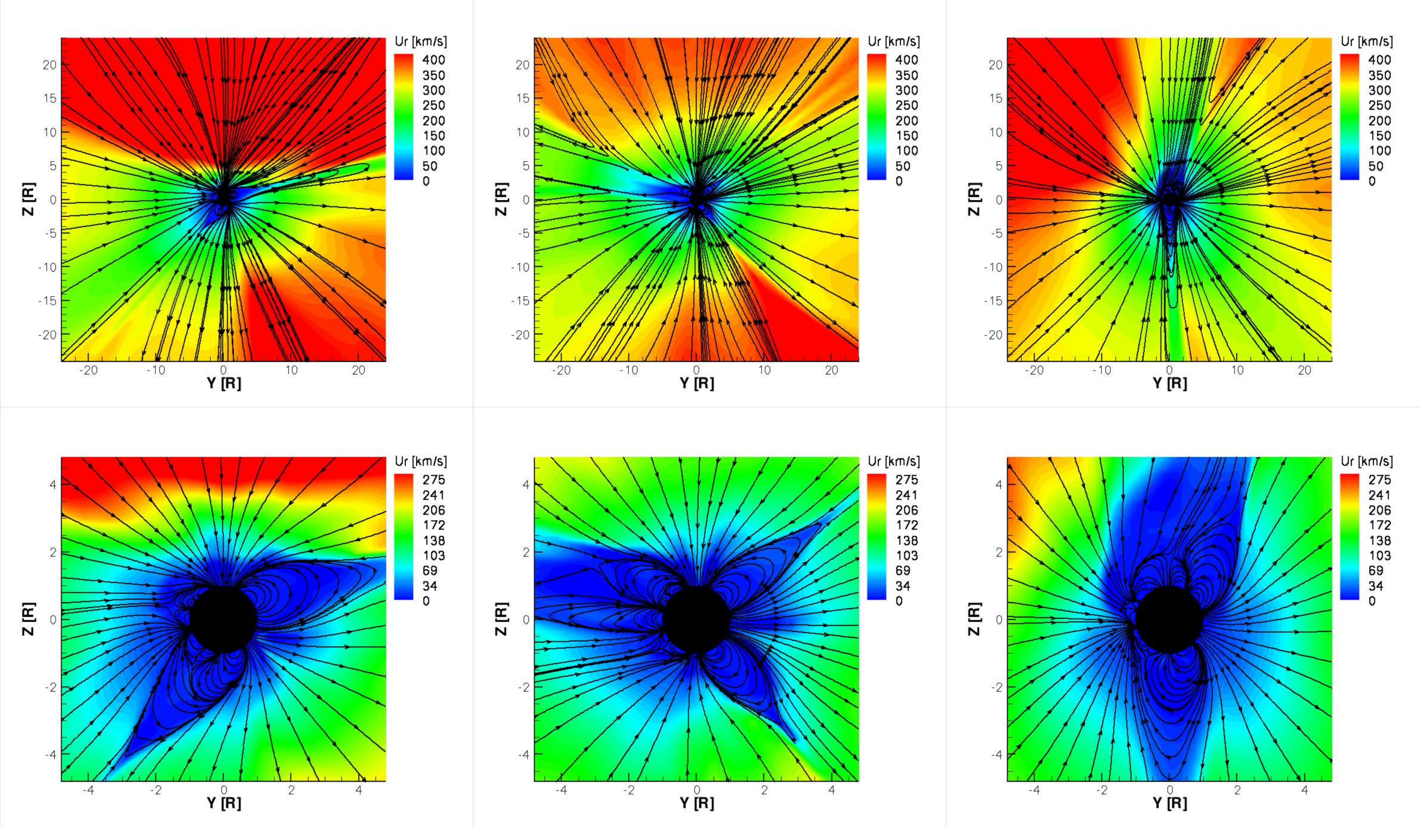} 
\caption{Top: MHD wind solution, displayed on a meridional cut, driven
  by the low-resolution magnetogram (left), by the processed map (middle), and by the
  high-resolution MDI magnetogram (right). Bottom: zoomed in images of
  the same solutions}
\label{fig:WindSolutions}
\end{figure*}

Using BATS-R-US we simulate the MHD solar wind for three different
magnetograms.  The WSO low-resolution modeled map (shown in the bottom left plot of
Figure~\ref{fig:magnetograms}), the processed map that results from
applying twice the HREM algorithm and boosting the parameter $ a $ to
$ 10 $ (shown in the bottom right plot of
Figure~\ref{fig:magnetograms}), and the high-resolution MDI
magnetogram (shown in the top right plot of the same Figure).  It is
worth to point out that, while these magnetograms have different length scales, the runs were
performed using the same magnetogram and MHD grid resolution.  The
wind solutions for these three magnetograms are shown in the top panel of
Figure~\ref{fig:WindSolutions} (from left to right respectively). 

It is clear that the loop structure changes
somewhat when artificially increasing the magnetogram spatial
resolution. The streamer near 70 degrees latitude (measured
clockwise from the north pole) 
of the low-resolution driven solution moves slightly towards the north
pole, while the one corresponding to a latitude close to 200 degrees moves counterclockwise about 90 degrees.
From the zoomed in plots of the same solutions (bottom panel of
Figure~\ref{fig:WindSolutions}), the shape of the loops near the solar surface
can be analyzed in more detail.  These show that the latter two
streamers also get narrower when artificially increasing the
resolution of the magnetogram. One can also see from these close in plots that the small double
streamer at latitude 270 degrees
becomes much larger, and the double streamer at near 200 degrees becomes a
simple streamer for the new solution.

\begin{figure*}
\centering
\includegraphics[width=\textwidth]{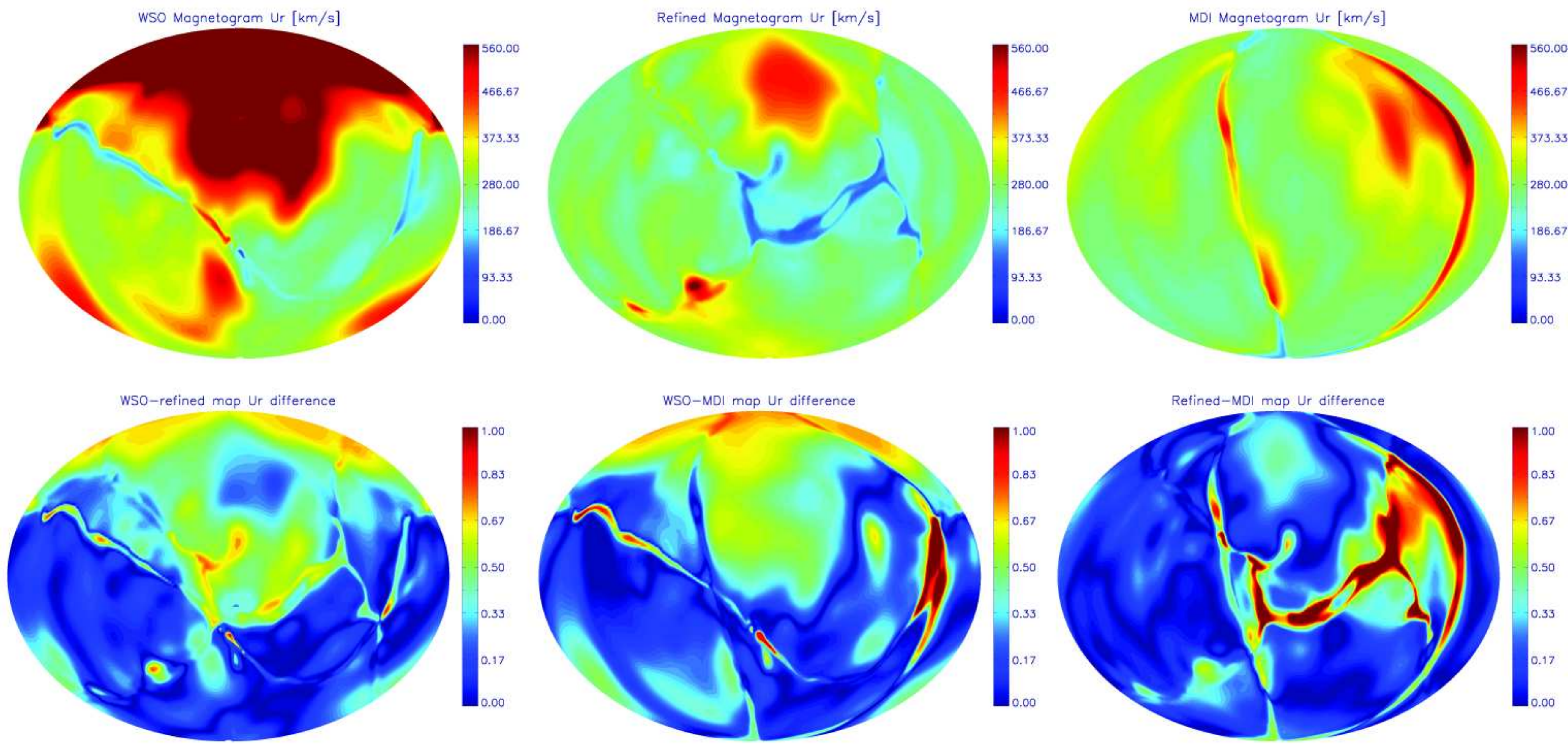} 
\caption{Projection maps of the
  radial speed for the WSO, the processed map and the MDI driven wind solutions extracted at $ r= 15 R_{sun}$ (top), and the
  normalized difference between them (bottom).}
\label{fig:projectionmaps_ur}
\end{figure*}

\begin{figure*}
\centering
\includegraphics[width=\textwidth]{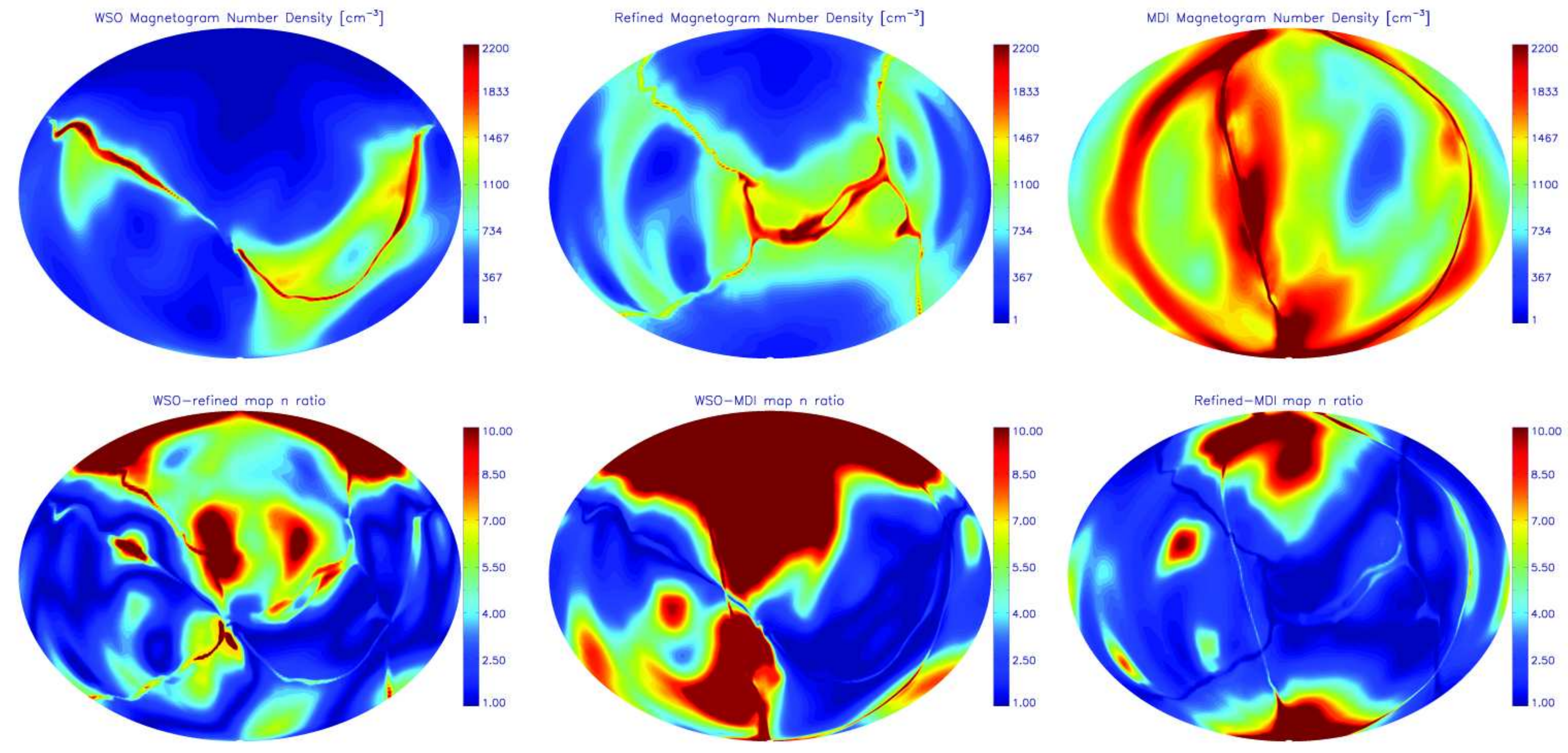} 
\caption{Projection maps of the
  number density for the WSO, the processed map and the MDI driven wind solutions extracted at $ r= 15 R_{sun}$ (top), and the
  normalized difference between them (bottom).}
\label{fig:projectionmaps_n}
\end{figure*}

For a more quantitative analysis we compare the radial speeds and number
densities of the different wind solutions at $ 15
R_\odot$. Figure~\ref{fig:projectionmaps_ur} shows the projection maps
of the wind radial speeds at $15 R_\odot$ for the WSO, the processed map
and the MDI driven solutions (top panel), and the normalized
difference between them (bottom
panel). Figure~\ref{fig:projectionmaps_n} shows the same projection
maps for the number densities at the same radial distance. 
The difference maps are absolute
differences normalized to the absolute magnitude of the first one:  $abs(u_1-u_2)/abs(u_1)$. And the ratio is the largest of the
ratios between two densities:  $max(n_1/n_2, n_2/n_1)$.
From the WSO-refined map difference (bottom left plots of both Figures) it is clear that the wind
solution changes significantly when introducing fine structure in a consistent
way.  The wind speed solutions compare better with the MDI solution when the
simulations are driven by the processed magnetogram. This is particularly clear from the bottom left plot of Figure~\ref{fig:projectionmaps_ur}
when comparing the wind speeds
at high latitudes for the low-resolution based solution (left top plot) and the one driven by 
the processed map (middle top plot), with the MDI driven wind
solution (right top plot). In the first case, very fast winds of about 500 km/s
reach the $15 R_\odot$ surface, and even closer to the solar surface
(see also top
left plot of Figure~\ref{fig:WindSolutions}), while this is not the case
for the MDI driven solution. This discrepancy is suppressed for the processed 
magnetic map (top middle plot of Figure~\ref{fig:projectionmaps_ur}
and top middle plot of Figure~\ref{fig:WindSolutions}).

\noindent Interestingly, one can also see that the differences in the refined-MDI plots
are much more local than the WSO-MDI. We expect the refined and the
MDI driven solutions to be different due to differences in
the magnetograms themselves. However, the fact that these differences become more
local suggests that the enhanced map performs better when
simulating the wind solution.
 

\subsection{X-Ray emission}
 
\begin{figure*}
\centering
\includegraphics[width=4in]{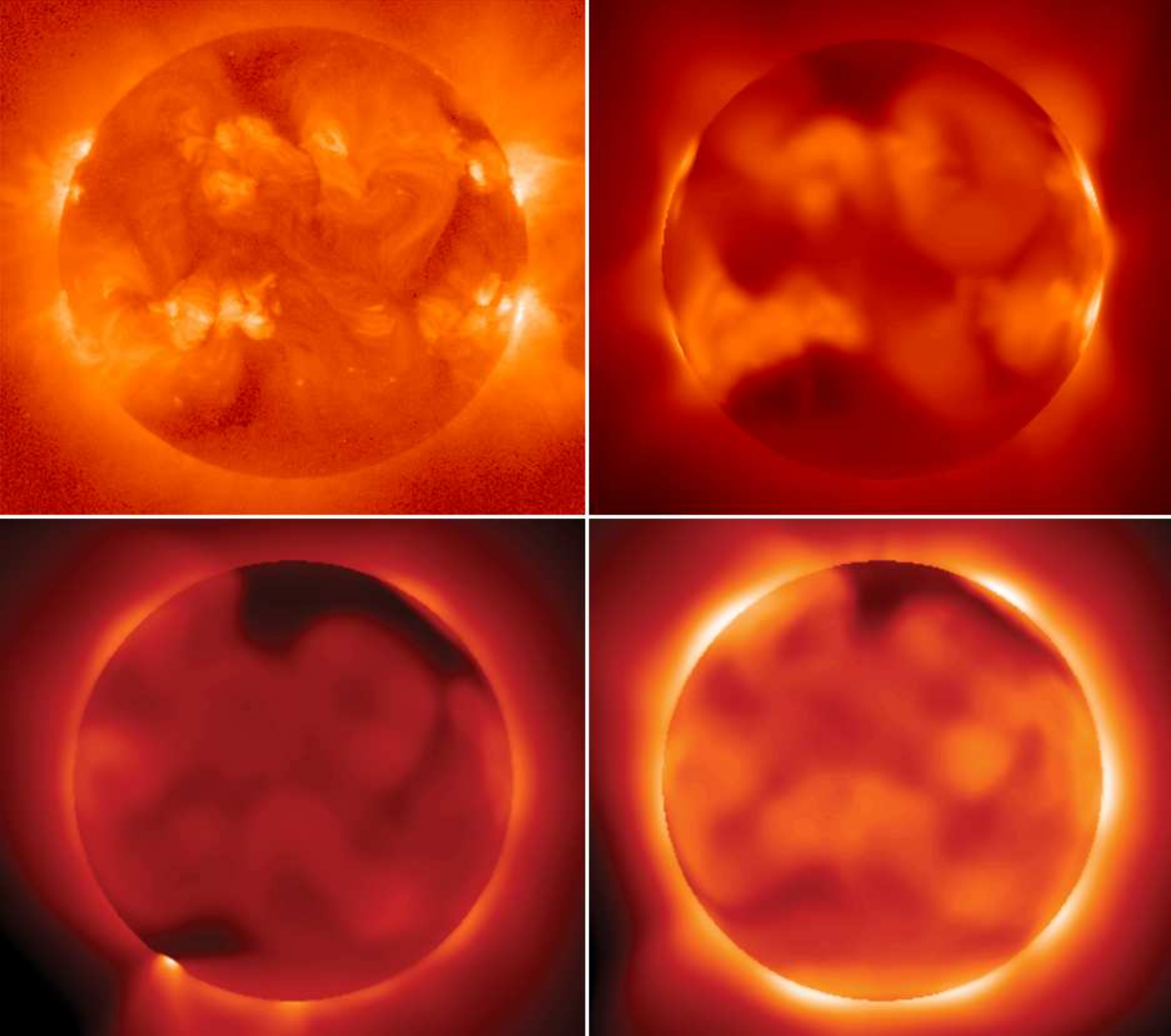}
\caption{A
  real X-ray image obtained by {\it Yohkoh} contemporaneously with the magnetograms  and corresponding to the model viewing angle (top left panel).
 X-ray emission modeled for the high-resolution MDI map (top right), for the low-resolution magnetogram (bottom left), and
 for the processed map (bottom right).}
\label{fig:SXR_all}
\end{figure*}

Corona models for the three magnetograms are compared with the X-ray morphology observed by
{\it Yohkoh} in Figure~\ref{fig:SXR_all}. Each image is reproduced on a logarithmic scale to provide the dynamic range to see fainter as well as very bright regions. The model
driven by the MDI map reproduces very well the bright active regions and some of the loop arcades of the {\it Yohkoh}  X-ray image. Fine and
fainter details are lost, but overall the synthetic image produces a
reasonable likeness of the observed one. The low-resolution WSO map 
solution captures some of the gross active region emission, but fails to separate the active regions near the western limb and loses nearly all the detail, including the active region near the eastern limb at mid-southerly latitude.

A significant amount of structure arises when the
resolution of the magnetogram is artificially increased. Many active 
regions brighten up and start to resolve.
In particular, two active regions in the southern hemisphere (one near the
center of the equator and
 the other near the western limb at mid-southerly latitude), two along the
 equator (one near each limb), and one at high latitudes near the
 eastern limb, become much more noticeable. 

The values of the X-ray fluxes in the images of
Figure~\ref{fig:SXR_all} are: $2\cdot 10^{25}\;ergs\;s^{-1}$ for the WSO
solution (bottom left plot), $5\cdot 10^{25}\;ergs\;s^{-1}$ for the
processed map solution (bottom right plot), and $1\cdot 10^{27}\;ergs\;s^{-1}$ for the MDI
solution (top right plot).  The X-ray flux increases $150\%$
when introducing fine-scale structure.  Although this increase is
rather small as compared to the gap between the WSO and MDI X-ray
fluxes, it is clearly in the right direction.


\section{DISCUSSION}
\label{sec:Discussion}

Our MHD simulations reveal that both the modeled magnetic topology and X-ray
morphology are affected by the fine-scale surface
magnetic structure.

The wind solution changes
significantly when using the MDI solution.  The reason for this is that the wind models depend strongly on the
expansion factor, defined in Section~\ref{sec:Model}.  Adding small-scale structure
to the magnetogram introduces a new mapping of the magnetic field that
changes the distribution of the expansion factors.

When modeling the 
X-ray emission that originates close to the solar (or stellar) surface, the small-scale magnetic topology of the 
surface is important. The reason for this is that the field that dominates the closed loop structure and contains the brightest hot, X-ray emitting plasma tends to be relatively small-scaled, and including finer detail can dramatically change the loop structure
at low altitude.  The 
high-resolution MDI magnetogram includes areas where there is mixing of
polarity.  The regions of mixed polarity give rise to the small loops which contribute substantially to
the X-ray emission.  In the low-resolution version of
the magnetogram, these areas  are smeared.  The mixed polarity regions
around the boundaries between larger regions of opposite polarity are
lost, and only the large dominant regions of polarity remain.  Consequently, the
small-scale loop population is lost.  In general, then, the
low-resolution solution includes a smaller number of larger loops,
while in reality the surface is covered by larger number of smaller
loops. The large loops obtained by the low-resolution map are not hot
and dense enough to compensate for the reduction of simulated X-ray
flux from the missing population of smaller low-lying loops.  Thus, it is not surprising that the X-ray
model solution shows a lot more structure when the magnetogram
resolution is artificially increased.  
There is a slight increase in the X-ray flux for the refined map, but it is rather small. This is probably due to the fact that the refinement does not introduce a significant amount of new closed loops that can contribute to the X-ray flux.

It is observed in solar active regions that areas of one dominant
polarity actually comprise bundles of smaller more discrete regions of
a dominant polarity with more spatially concentrated magnetic field. The fact that both wind and X-ray morphology solutions show changes when the spatial resolution of the magnetogram is 
artificially increased suggests that further work is needed to
decide whether introducing more mixed polarity
(probably hidden within each monopole region) makes a bigger
difference.   Is the artificially enhanced map more ``realistic'' than the WSO one?  It is tempting to describe the former as better representing the observed {\it Yokoh} image than the latter.  Lower resolution maps could then be made perhaps slightly more ``realistic'' by applying the deconvolution and enhancement techniques explored here to include more fine-scale mixed polarity.  

A cautionary note has then to be sounded regarding the reconstruction of coronal X-ray  emission from stellar ZDI magnetograms.   We remarked in Sect.~\ref{sec:Magnetograms} that stellar magnetograms were generally of a much lower spatial resolution than even the WSO magnetograms.  For a star with a complex surface magnetic field distribution, like the active Sun, reconstructed X-ray emission is unlikely to bear a striking resemblance to reality.  For the case of very active stars like AB~Dor, the situation is less clear.  Near the equator, the  resolution of AB~Dor magnetograms is of a similar order to that of the WSO ones.   The key question for the propriety of reconstructed X-ray images regards the true complexity of the surface magnetic field and whether there is a lot of unresolved structure.  

\citet{Johnstone10} and \cite{Arzoumanian11} note that, since the
Zeeman signature is suppressed in the dark spotted regions of the
stellar surface, ZDI magnetograms are censored in that they do not
reconstruct reliably the field in star spots.  \cite{Arzoumanian11}
found that artificially including field with randomly oriented
polarity in spotted regions lead to significantly different potential
field reconstructions of the coronal morphology and X-ray rotational
modulation for AB~Dor, but less so for the mostly dipole-dominated M
dwarf V374 Peg.  \citet{Johnstone10} also studied the effect of limited spatial resolution using synthetic maps with initial resolution of slightly more than $1\deg$---similar to  the WSO maps studied here---and smearing these to latitudinal and longitudinal resolutions of $11\deg$ and $8\deg$ (the latter at the equator).  They found that limited spatial resolution does not have a large effect on predicted emission measure for potential field coronal models.  Their study covers a much lower resolution regime than we study here though, and the limited sensitivity to spatial resolution is likely partly related to the limited degree of structure at very small spatial scales in their synthetic maps.    The existence of compact structures with high plasma densities in active stellar coronae \citep[e.g.][]{Testa04,Ness02} hints at small-scale magnetic structure.  Further investigation of finite resolution effects on the most active stars would be well-motivated.


\section{SUMMARY and CONCLUSIONS}
\label{sec:Conclusions}

Low resolution maps for the magnetic structure of the surface of stars
have been widely used to analyze different stellar properties under
the assumption that these maps were representative enough for the
purposes addressed. This assumption has already been subject to some scrutiny.   
Using a set of MHD-based simulations of the solar corona and wind, we
have investigated how the spatial resolution on scales of $5\deg$ and
smaller of the input magnetogram can change the derived structure and
morphology of the wind, the magnetic field, and the coronal X-ray
emission.  We conclude that both wind properties and X-ray
morphology are significantly affected by 
the fine details of the surface magnetic field.  The
fine-scale structure of the fiducial magnetogram has an impact on the flux tube sizes
and expansion factors, therefore affecting the wind solution.  Similarly, when
modeling the X-ray emission, it becomes crucial
to take into account the very small-scale stellar magnetic
structure, because it dominates the X-ray emitting loop
population.  The regions of opposite polarity that support closed loops need to be reasonably well-resolved in
order to obtain a realistic X-ray morphology.  We find that for both the MHD model wind and the
low corona X-ray model, the solution changes significantly when artificially increasing
the resolution of the magnetogram by dividing magnetic poles into bundles of stronger but more compact magnetic flux.  This fake enhancement also arguably results in perhaps slightly more ``realistic'' synthetic coronal X-ray images.


\acknowledgments

We thank the unknown referee for the valuable comments on our
manuscript.
OC is supported by NASA's CXC contract TM2-13001X.
JJD was supported by NASA contract NAS8-39073 to the {\it Chandra X-ray Center}, and thanks the Director, H.~Tananbaum, and the CXC science team for advice and support. 
Simulation results were obtained using the Space Weather Modeling
Framework, developed by the Center for Space Environment Modeling, at the University of Michigan with funding
support from NASA ESS, NASA ESTO-CT, NSF KDI, and DoD MURI.






\end{document}